\documentclass{aa}

\usepackage{graphicx}
\usepackage{txfonts}

\begin{document}

\title{Assessment of different formation scenarios for the ring system of (10199) Chariklo}

\author{Melita, M.D.
\inst{1,2}
\and
Duffard, R.
\inst{3}
\and
Ortiz, J.L. 
\inst{4}
\and
Campo-Bagatin, A. 
\inst{4,5} }

\institute{Instituto de Astronom\'ia y F\'isica del Espacio 
(CONICET-UBA), CABA, Argentina\\
\email{melita@iafe.uba.ar}
\and
Facultad de Ciencias Astron\'omicas y Geof\'isicas, Universidad
Nacional de La Plata, Argentina
\and
Instituto de Astrof\'isica de Andaluc\'ia, IAA-CSIC, Granada, Spain
\and
Departamento de F\'isica, Ingenier\'ia de Sistemas y Teor\'ia de la
Se\~nal, Escuela Polit\'ecnica Superior, Universidad de Alicante,
Spain
\and
Instituto de F\'isica Aplicada a las Ciencias y la Tecnolog\'ia,
Universidad de Alicante, Spain
}

\abstract{The discovery that the centaur (10199) Chariklo 
possesses a ring system opens questions about their origin.}%
{We here asses the plausibility of different scenarios for the origin 
of the observed ring system. }%
{We first consider the possibility that the material of the ring 
originated in the disruption of a satellite that had reached a critical
distance from the centaur. We discuss the conditions for the putative
satellite to approach the centaur as a consequence of tidal
interaction.  A three-body encounter is also considered as a transport
mechanism. In addition, we study the case in which the ring is formed
by the ejecta of a cratering collision on the centaur and we
constrain the collision parameters and the size of the resulting
crater of the event. Finally, we
consider that the ring material originates from a catastrophic
collision between a background object and a satellite located at a
distance corresponding to the the current location of the ring. 
We compute the typical timescales for these scenarios.  }%
{We estimate that in order to be tidally disrupted a satellite
would have had to be larger than approximately $6.5\ km$ at the location
of the rings. However the tidal interaction is rather weak for objects
of the size of outer solar system bodies at the ring location,
therefore we considered other more effective mechanisms by which a
satellite migt have approached the centaur.  Collisonal scenarios are both
physically plausible for the formation, but semianalytical
estimations indicate that the probability of the
corresponding collisions is low under current conditions.}%
{}%
%


\maketitle

\section{Introduction}

The remarkable discovery of a ring system around (10199) Chariklo
\citep{Braga2014}, naturally raises a number of questions about their  
origin and the formation environment of small bodies in the outer
solar system, since  rings have been known so far  to exist only around
the four largest planets.

(10199) Chariklo is the largest known centaur, similar in
size to 2002 GZ32 \citep{Duffard2014b}. The centaur
orbit is between those of Jupiter and Neptune, its semimajor
axis is $15.8\ AU$ and its current eccentricity is
$0.175$, with an effective radius of approximately $ 124\ km \pm 9\ km$,
\citep{Duffard2014} and a geometric albedo of $0.035\pm 0.011$
\citep{Fornasier2013}.  

The ring system has two components of widths $3\ km$ and $7\ km$,
separated by a gap of about $8\ km$  and they are located between $391\ km$
and $404\ km$ from the centaur and their estimated mass is
equivalent to that of a body of $1\ 
km$ radius with density of $1\ g\cdot cm^{-3}$ \citep{Braga2014}. 

It has been concluded
\citep{Duffard2014} that the dimming of the centaur between 1997 and
2008 and the gradual decreasing of water ice features from the
spectra are due to the changing view geometry of the rings. This
implies that the rings are primarily composed of water ice. The
steepness of the ocultation profile at the edges of the rings, the
existence of the gap, as well as their short estimated spreading
lifetime of only $10^4 yr$ strongly suggest that the rings are
confined by kilometer-sized shepherd satellites \citep{Braga2014}.  

Further observations \citep{Sicardy2014} revealed that the rings may
have an azimuthal structure, that can be attributed to an eccentric
shape of the rings, which implies that the orbits of the ring
particles are precessing rigidly. Using a model where differential
precession is compensated for by the self gravity of the ring
\citep{Chiang2000}, \citet{Pan2016} have estimated the mass of the
rings. Nevertheless, it has been put forward that shepherd satellites
do play a fundamental role in keeping apsidal alignment
\citep{Papaloizou2005} and therefore this mass estimate may need to
be revised. Furthermore, it is difficult to envisage the origin of the
shepherd satellites in a formation scenario driven by cometary-like
activity, as put forward by those authors.

It has also been suggested that the rings originated in the disruption
of a parental centaur during a close encounter with one of the
giant planets \citep{Hyodo2016}, which should have taken place at
about 2 planetary radii. This scenario can  successfully explain 
not only the dynamical origin of the rings, but also their
composition, since the material is extracted from the ice-rich mantle
of the incoming body. 

On the other hand, detailed dynamical simulations show that the centaur
ring system is stable \citep{Araujo2016} under close encounters
with the giant planets that (10199) Chariklo experienced in its past
history, which did not occur at distances smaller that
about 10 planetary radii. 

In the following sections, we discuss several possible formation
mechanisms that might explain the presence of the rings. We have studied
the conditions to disrupt a satellite in the neighborhood of the
centaur, as well as different orbital transport mechanisms to that
location. We have also considered the possibility that the rings have
originated in either a cratering collision on the 
centaur, or in a catastrophic collision on a satellite located where the
rings are, and we have also estimated the probability of these
events in the lifetime of typical centaurs. In the last two
sections we discuss our results and lay out our conclusions.  

\section{Roche limit of (10199) Chariklo} 

For a spherical satellite, the classical Roche limit, $d$, in which
the self-gravity of the satellite is equal to the tidal force of the
primary (the centaur, in our case) is given by 
\citep[see for example][]{Murray1999}

\begin{equation}
d = R_s \left (3\frac{M}{m} \right )^{1/3} ,
\label{eq:1}
\end{equation}

where $R_s$ is the radius of the satellite, $M$ is the mass of the
centaur and $m$ the mass of the satellite. 

It is interesting to compare the value 
of the Roche limit  $d$ with the so called Roche
zone or Hill's radius, $d_H$, which is the distance to the corresponding $L_1$ and
$L_2$ Lagrange equilibrium points in the restricted three-body problem, 
\begin{equation}
d_H = a \left (frac{M}{3 m} \right )^{1/3} ,
\label{eq:1}
\end{equation}
where $a$ is the semimajor axis of the orbit of the satellite. In the
case of (10199) Chariklo is approximately $d_H \approx 2.5\ 10^{5} km$.

We note that the shape of (10199) Chariklo seems to be very close to that of an
ellipsoid of revolution with semiaxes  $A=B=117\ km$ and $C=122\ km$
\citep{Braga2014, Duffard2014}. Therefore its mass can be approximated
as:
$$M = \frac{4\ \pi}{3} \rho A^2 C , \label{eq:M}$$
where $ \rho $ is the unknown density of Chariklo.
If we assume that the satellite is also a symmetric ellipsoid of
revolution with axes $A_s=B_s$ and $C_s$, its mass,
$m$,  can be written as:
$$m = \rho_s A_s^3 \epsilon    $$
where $ \rho_s $ is the density of the satellite and 
$\epsilon$ is defined as:
$$ \epsilon = \frac{4\ \pi}{3} \gamma_s ,$$
where the sphericity factor, $\gamma_s$, is:
$$ \gamma_s = \frac{C_s}{A_s}.$$
Values of $\gamma_s$ between $0.1$ and $10.$ can be considered
realistic, as these values correspond to lightcurve amplitudes of $0.4
mag's$, which is the largest value observed for fast-rotator
asteroids \citep{Warneretal2009}.  



Hence, to account for the non spherical shape of both bodies, we write:
\begin{equation}
d = \left(3 \frac{\delta}{\gamma_s} {A^2 C} \right)^{1/3},
\label{eq:2}
\end{equation}
where $\delta = \rho/\rho_s$.

\begin{figure}[p!]
\centering
\includegraphics[angle=-90,width=\textwidth]{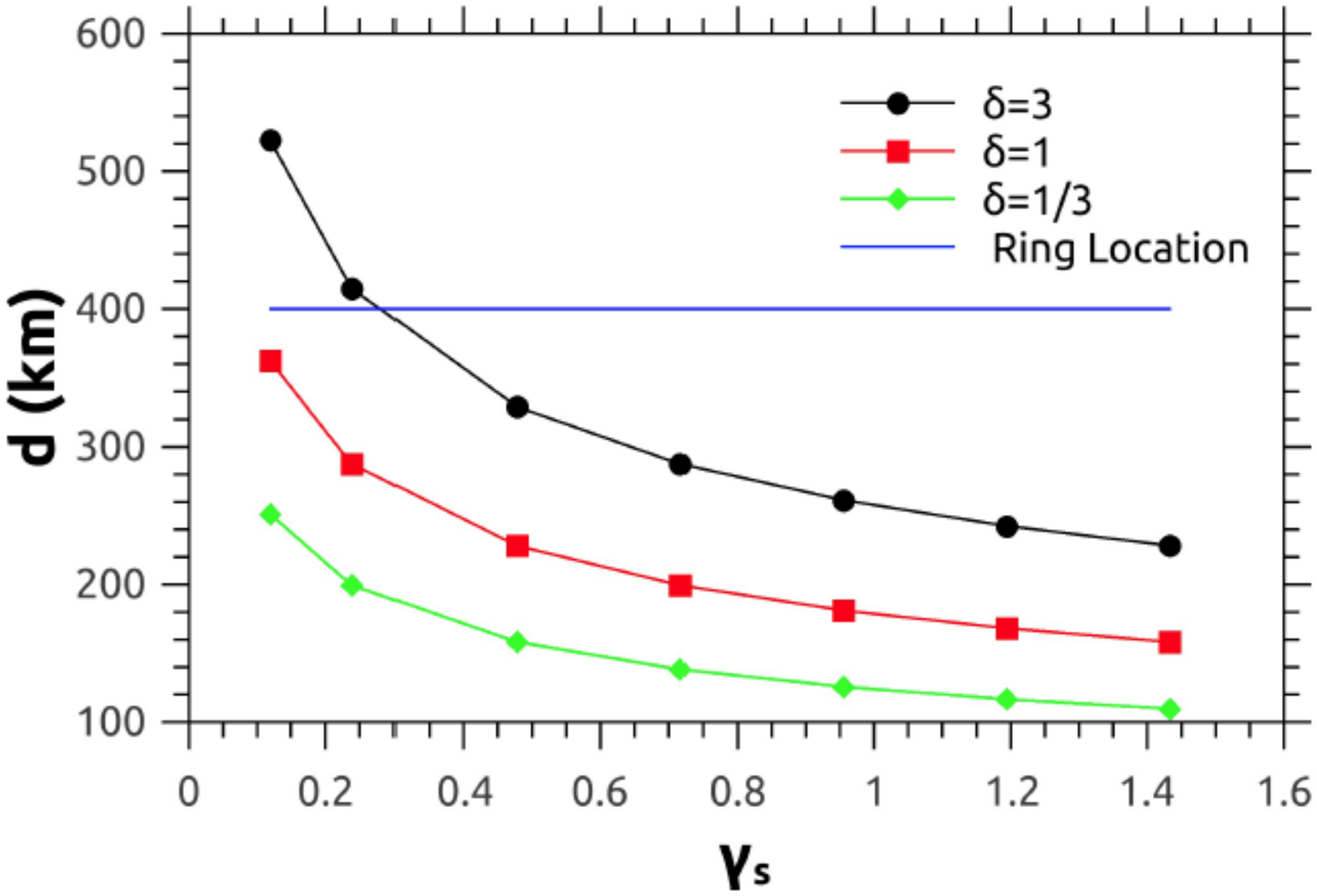}
\caption{Location of the Classical Roche limit, $d$, for (10199) Chariklo
as a primary as a function of
$\gamma_s$ for various values of $\delta$ (see 
text for details). 
}
\label{fig:1}
\end{figure}

In Figure \ref{fig:1}
we plot the values of $d$ (the Roche limit distance), for (10199)
Chariklo as a primary, as a
function of $\gamma_s$ for various values of $\delta$. Note that the
values of  $d$ are asymptotic for values of $\gamma_s$ larger than
$\approx 1.6$. For
$d$ to be in the range where the rings are observed today, both those
parameters take remarkably extreme values.

\subsection{Disruption distance}
\label{sec:DD}

Now we shall also consider the fact that the satellite material 
most probably possesses a non negligible tensile and/or shear strength, in order to estimate
the size that an object can posses without being tidally disrupted.
We work out the splitting distance, $\Delta$, depending on the strength
regime and its rotational state \citep{Dobrovolskis1990}. 
We consider a spherical satellite
with radius $R_s$ and density $\rho_s$ orbiting (10199) Chariklo,
with mean radius $R$ and density $\rho$. Depending on the pressure at
the center of the body, given by

$$p_0 = 2/3\ \pi G \rho_s^2 R_s, $$ 

the {\it Weak} regime occurs when the tensile strength, $T$, is in the
regime $T<p_0$, and the {\it Strong} regime when $T>p_0$.  We assume a
value of the tensile strength, $T=0.88\ \rho_s$ \citep{Davidsson1999}
(density must be in I.S units, to get $T$ in $Pa$). We follow
\citet{Dobrovolskis1990} and, since we are interested in the low
limits of distance $d$, we assume that the body is not tidally locked.
In that case the disruption distance for icy materials is given in
Table~\ref{tab:2}, where we have assumed that the shear-strength of
ice, $S$ follows the approximate expression: $S=3T$
\citep{Dobrovolskis1990}. 
 
\begin{table}
\begin{tabular}{|l|c|}
\hline
Regime & $Delta$ \\ \hline  
Weak & $1.37\ R\ \left( \frac{\rho}{\rho_s} \right)^{1/3}\
\left( \frac{p_0}{S}\right)^{1/3}$ \\ \hline 
Strong & $1.19\ R\ \left( \frac {\rho}{\rho_s} \right)^{1/3}\
\left( 1+\frac{T}{p_0} \right)^{-1/3}$  \\ \hline 
\end{tabular}
\caption{Disruption distance, $\Delta$, according to
\citet{Dobrovolskis1990}, for not tidally locked icy bodies. 
} \label{tab:2}
\end{table}

Figure \ref{fig:3} shows the values of the splitting distance,
$\Delta$, for the two different strength regimes described above, as a
function of the satellite radius, in a case in which the
satellite and the centaur densities are both equal to $1\ g\cdot
cm^{-3}$, where we note that
the radius of an object has to be larger than about
$6.5\ km$ to be broken up at the location of the rings of (10199)
Chariklo.  

The validity of this result can be checked against the results of
\citet{HolsapleMichel2008} who computed limiting distances
considering material strength, parameterized at different friction
angles as well as tidal
and centrifugal forces. We deduce from Figure 4 in that article, that
if the disageggated satellite
is $R_s \approx 3\ km$ and 
is tidally locked, i.e. its intrinsic rotation period is equal to its
orbital period, then the limiting distance,
$d'$, is independent of the friction angle and
is given by: 
$$\frac{d'}{R_{CH}}\ \times \left (\frac{\rho_s}{\rho} \right)^{1/3}  \approx 1 ,$$
where $R_{CH}$ and $\rho$ are the mean radius and the density of the 
of the centaur. For a satellite with no intrinsic rotation, this
condition occurs
 for a physical radius of $R_s \approx 2\ km$, i.e. for this
 satellite's size the splitting
 distance is located at one centaur's radii when both densities are
 equal 
\citep[Figure 6, ][]{HolsapleMichel2008}. For $d'$ to be located at
the current location of the rings, the
ratio of the densities should take an unrealistic value of,
$$ \frac{\rho_s}{\rho} \approx \frac{1}{40}.$$

On the other hand, we also deduce that an object with a similar
density to that of the centaur would have to have a radii in the range
$4\ km < R_s < 7\ km$ and low-friction angle to disagreggate at $d' \approx 3.6 \times R_{CH}$,
where the rings are located at present 
\citep[Figures 4 and 6, ][]{HolsapleMichel2008}, which confirms the
result obtained with the model of \citet{Dobrovolskis1990}.

\section{Tidal effects}

In this section we discuss the orbital decay that is due to the tidal
interaction produced by a body  of the size of (10199) Chariklo on a
putative satellite.  

We apply the Darwin-MacDonald approximation for the tidal potential,
which only considers the terms with the lowest degree on $\cos{ \psi}
$ and $R_s/r$, where $\psi$ is the angle formed by the position
vector, ${\bf r}$, and the $x-y$ plane. We also assume the classical
expression of lag-angle subtended at the center of mass of the centaur
between the direction of its tide-raised bulge and the longitude of
the satellite, $l$, that is  $sin(l) = 1/Q$, where $Q$ is the quality
factor of the body that raises the tide. Hence, the rate of change of
the semimajor axis of the satellite that is due to the tides raised by the
centaur, can be estimated following \citep{Murray1999}

\begin{equation}
\dot{a} \approx sign(\omega_{CH}-n)\ \frac{3 k_2}{2 \alpha Q_{CH}}\ 
a\cdot  n\ \frac{m}{M}\  \left( \frac{R_{CH}}{a} \right)^5.
\label{eq:4}
\end{equation}

Here, we assume that the spin of (10199) Chariklo and the
satellite have the same sense;  $a$ is the semimajor axis of the
orbit of the satellite, $n=\sqrt{GM/a}$ is the local orbital frequency
in the satellite-centaur two-body problem, $R_{CH}$ is the mean radius
of the centaur, $\omega_{CH}$ its intrinsic rotation frequency,
$Q_{CH}$ its quality factor and $k_2$ its Love number. We estimate the
value of $k_2$ as \citep{Murray1999}:

$$k_2 = \frac{1.5}{( 1 + \bar{\mu} )}.$$

The effective rigidity $\bar{\mu}$ of the object is estimated as
\citep{Murray1999}:

$$\bar{\mu} = \frac{19 \mu }{\left[ 2 \rho g(A) A\right] }, $$

where $\mu$ is the mean rigidity of the material and $g(A)$ the
acceleration of gravity on its surface \citep{Murray1999}. We choose
$\mu = 4\ N\cdot m^{-2}$, which is typical of an icy body, and we obtain $k_2=
1.6\cdot 10{-4} $, for $\rho = 1\ g\cdot cm^{-3}$. We assume a typical
value for icy solar system minor bodies of $Q_{CH}=100$
\citep[Table 4.1][]{Murray1999}. 

>From Equation \ref{eq:4}, we estimate the time of travel $\tau_{12}$
between two semimajor axes $a_2$ and $a_1$, that is due to the
tides raised on the centaur, as

\begin{equation}
\tau_{1 2} = sign(\omega_{CH} - n)\ \frac{2}{13}\ C_T\ \left(a_2^{13/2} - a_1^{13/2} \right),
\label{eq:5}
\end{equation}

where 
$$C_T=  \frac{2}{9}\  \left( \frac{Q}{k_2} \right)\ \left( \frac{M}{m} \right)\
\left( \sqrt{\frac{1}{R_{CH}^5 G M}} \right).$$ 

For orbital decay to occur, the satellite must be inside a synchronous
orbit. On one hand, using the quoted parameters of the centaur, the
orbital period of the ring particles of (10199) Chariklo is
approximately $20 hs$ (see Figure \ref{fig:sk0}), which means that, for the
orbit to evolve inward the rotational period of (10199) Chariklo
should be --or should have been at the time of the formation of the
rings-- longer than that value. On the other hand, a shorter intrinsic
rotational period of about $7 hs$, corresponding to a double-picked
light curve, has been determined for the centaur by
\citet{Fornasier2013}, using SOAR (4m) data from five non-consecutive
nights. The synchronous orbit would then be located at about $676 km$
from the centaur, i.e. beyond the location of the rings.  Therefore,
the orbital evolution of a hypothetical satellite located where the
rings are observed at present, would tend to expand its orbit. 

We nevertheless consider the case in which the rotation rate of
the centaur may have been slower when the tidal interaction with the
satellite started. Taking the final distance, $a_2$, as
the mean location where the rings are observed today, we compute the
timescale necessary to reach this distance as a consequence of the semimajor
axis decay due to the tidal interaction with the centaur.
In Figure \ref{fig:5} we plot this timescale as a function of the
starting distance (which we assume to be larger than $500\ km$) for different
values of the radius of the satellite and its density. This timescale
is shorter than the age of the solar system if the radius of the
satellite is larger than $10\ km$. For the semimajor axis change to
have a considerable evolution in the typical dynamical lifetime of a
centaur, we find that the satellite would have had to be larger than $200\ km$.  

We also consider the orbital evolution caused
by the tides raised by the centaur on a hypothetical satellite.
When we applying the same approximation as above, the semimajor axis change
is estimated as

\begin{equation}
\dot{a} \approx sign(\omega_{S}-n) \frac{3 k_2(S)}{2 \alpha Q_{S}}
a\cdot n\ \frac{M}{m} \left( \frac{R_{S}}{a}\right)^5,
\label{eq:4b}
\end{equation}

where $\omega_{S}$ is the intrinsic rotation rate of the satellite,
$R_{S}$ its physical radius, $Q_{S}=Q_{CH}=100$ its quality factor and
$k_2(S)$, its Love number, which is estimated in a similar way as
previously described for $k_2$, where satellite parameters are used. 
Now, we estimate the time of travel 
, $\tau'_{12}$ between two
semimajor axes locations $a_2$ and $a_1$, due to the tides raised on
the satellite as:

\begin{equation}
\tau'_{1 2} = sign(\omega_{S} - n)\ \frac{2}{13}\ C'_T\
\left(a_2^{13/2} - a_1^{13/2} \right),
\label{eq:5b}
\end{equation}

where

$$C'_T=  \frac{2}{9}\  \left[ \frac{Q_S}{k_2(S)} \right]\ \left(
\frac{m}{M}
\right)\
\left( \sqrt{\frac{1}{R_S^5 G M}} \right).$$

In Figure \ref{fig:5b} we plot $\tau'_{1 2}$ as a function of the initial
orbital semimajor axis for different sizes of the satellite. We
assume that $\omega_{S} < n$ such that the orbital evolution occurs
inward.  The satellite density is assumed to be equal to $1\ g\cdot 
cm^{-3}$, but we note that plausible changes of this parameter do
not alter the conclusion substantially. The values of 
$\tau_{1 2}$ are about one order of magnitude
higher than $\tau'_{1 2}$ and  on the order of the lifetime of a centaur if
its physical radius is larger than $100\ km$.  Therefore we conclude
that the tidal interaction is effective in bringing a satellite to a
distance to the centaur where it can be disrupted, only if the satellite
size is similar to that of the centaur itself and it is spinning
more slowly. 


\section{Three-body encounter}
\label{sec:3bp}

Tidal interaction is not the only mechanism capable of bringing a body
within the breakup distance to (10199) Chariklo. A three-body
encounter between a centaur and two field objects could be the cause
for one or both of them to reduce its relative energy with respect to the
centaur and become gravitationally bounded. The bounded object can
become the origin of the rings if its captured orbit 
occurs at a distance where tidal forces can
disagreggate it. To evaluate the plausibility of this mechanism, here we shall consider
a very simple model for the three-body encounter between a centaur of
mass, $M_{CH}$, and two field objects, $a$ and $b$ (see Figure
\ref{fig:sk1}, with masses $m_a$ and $m_b$ and radii, $R_a$ and $R_b$, all assumed to have
spherical shape and $1\ g\cdot cm^{-3}$ density. 

Neglecting the mutual gravitational energy between bodies $a$ and $b$, 
the energy before the collision, $E_1$, is given by:
\begin{equation}
E_1 = \frac{1}{2} m_a V_a^2 + \frac{1}{2} m_b (V_a +\Delta V_b)^2 -
\frac{G M_\odot m_a}{r_a} - \frac{G M_CH m_b}{r_b} + \frac{G
M_{CH} m_a}{r_{CH a}} + \frac{G M_{CH} m_b}{r_{CH b}} 
\end{equation}
where $r_a$ and $r_b$ are the heliocentric distances of bodies $a$ and
$b$, $r_{CH a}$ and $r_{CH b}$ and their distances with respect to the
centaur. While, the  energy after the collision, $E_2$, is,
\begin{equation}
E_2 = \frac{1}{2} (m_a
+ m_b)\cdot V_f^2 - \frac{G M_\odot (m_a + m_b)}{r} + \frac{G M_{CH}
(m_a + m_b)}{r_{CH}},
\end{equation}
and the conservation of the linear momentum, gives:
$$V_f = V_a + \frac{m_b}{m_a + m_b} \Delta V_b$$. 
The energy dissipated in the collision is $\Delta E = E_2 - E1$.

The gravitational energy changes with respect to the Sun and the centaur during the
collision are negligible because the distances are practically
unchanged. Therefore we can write:
$$\Delta E = \frac{1}{2} \frac{m_b^2}{m_a + m_b} \Delta V_b^2$$

The final orbit of the merged body is bounded to the
centaur at a final distance, $a_f$, therefore we can write, 
\begin{eqnarray}
\frac{G\ M_{CH}}{2 a_f} \left (m_a + m_b \right)^2 &=& \frac{1}{2}\ \Delta V_b^2\ m_b^2.
\label{eq:3bb}
\end{eqnarray}
If $a_f=a_{RINGS}$, the radii of the
bodies taking part into the collision are related by:
\begin{equation}
R_a^3 = R_b^3\  \left( \frac{ a_{RINGS}\ \Delta V_b^2}{ G\ M_{CH} } - 1 \right) 
\label{eq:3be}
\end{equation}

We find that, for this set of assumptions, the transport is
possibl when the largest body has a  radius of
$R_a= 6.5\ km$ and it is hit at a
relative velocity of $3\ km/s$ by a smaller body
of  radius $R_b = 330m$. Given its size, the largest body would
not disrupt, we estimated in section \ref{sec:DD}.

\section{Collision with the nucleus of (10199) Chariklo}

Our goal is to evaluate the possibility that the mass observed in the
rings of (10199) Chariklo, originated as ejecta from a 
collision on the surface of the centaur. We use the scaling expression
from \citet{Housen2011} to estimate the velocity, $v_e$, and the mass,
$m_e(v > v_e)$, of the material ejected with velocity $v > v_e$, as a
function of the distance to the center of the crater, $x$,
corresponding to an impact in the gravity regime:

\begin{equation}
m_e(v >  v_e) = \frac{3 k}{4 \pi} \frac{\rho}{\rho_s} \left [ \left (
\frac{x}{c} \right)^3 - n_1^3 \right ]
\label{eq:6a}
\end{equation}

\begin{equation}
v_e(x) = C_1\ v_r\  \left [ 
\frac{x}{A_i} \left ( \frac{\rho}{\rho_s} \right)^\nu \right]^{-1/\mu}\  
\left ( 1 - \frac{x}{n_2 R}  \right )^p,
\label{eq:6b}
\end{equation}

where $R = (A^2C)^{1/3}$, $\nu=0.4$, $k=0.3$, $p=0.3$, $n_1=1.2$,
$C_1=0.55$, and
$n_2=1.3$ are constants related to the icy nature of the body, and, $A_i$
is the radius of the impactor. 

We investigated which combination of distance to the center
of the crater, $x$,
physical radii and velocity of the
impactor would produce ejecta with velocities in a range such that
their final orbits would correspond to the ring system currently
observed, and which amount of ejected mass, $M_{ej}$, is set in orbit
in that location, that is,

\begin{eqnarray}
\frac{-G M}{2 a_o} <
\frac{-G M}{R} &+& \frac{1}{2} v_e(x)^2 < \frac{-G M}{2 a_i}  \\
M_{ej} \approx M_{RS}  &  \approx & M(v > v_i) - M(v > v_o)
\label{c1}\
\end{eqnarray}

where $a_o$ and $a_i$ are the currently observed limits of the ring
system, $v_o$ and $v_i$, the corresponding orbital velocities,
assuming circular orbits, $M(v > v_o)$ and $M(v > v_i)$ the mass
ejected with velocities higher than the corresponding limit values of
the ring system and $M_{RS}$ is the order of magnitude of the mass of
the ring. We also investigated how much ejected mass remains bounded
to the centaur in these cases.  

In Figure \ref{fig:vt} we plot of the combinations of 
impactor radii and velocity that satisfy the conditions set in
Eq. \ref{c1}, for each favorable pair, the corresponding values of ejected
mass, $M_{ej}$, crater radius and mean ejection distance from
it,  and $<x>$ is coded in color. The radius of the impactor is computed
assuming a density of 
$\rho_i=1\ g\cdot cm^{-3}$ and we find that values that render
favorable events are in a range between $200\ m$ and $1\ km$.

We also provide the corresponding plots for all the mass that remains
gravitationally bound to the centaur (right column). In all cases the
mass originates at about $1/3$ of the diameter of the crater, the
closest regions to the center of impact where the material is not compressed
into the target and it is able to escape. %
The mass that remains in bound orbits and the mass that falls back onto the surface of
the centaur is between 2 and 3 orders of magnitude larger than the
ejected mass that is estimated to remain in bound orbit at the
currently observed 
location of the rings. 

The past occurrence of such a collisional event could be easily
corroborated, if
the surface of (10199) Chariklo were observed at high resolution, because
the crater in which the rings would have had originated the rings
would be  in $20 -- 50\
km$ in diameter.

We can calculate the probability of such an event by scaling to Chariklo the number of
impacts computed  by \citet{Levison2000}, which give a value of 
$n_{U\&N} = 3.4 10^{-4} yr^{-1}$
for planets Neptune and Uranus, while we obtain $n_{CH} = 2.2 10^{-8} yr^{-1}.$

Therefore this event would be very unlikely at the current location of
the body,
given that the typical dynamical lifetime of a centaur is $\sim
10^{7}\ yr$ \citep{Horner2004}.  We note that if the collision rates in
the outer solar system were an order of magnitude higher, as derived
from recent observations of impacts on the giant planets
\citep{Hueso2013}, the probability of this event would be marginally
plausible over the dynamical lifetime of (10199) Chariklo as a
centaur. This object very likely originated in
the trans-Neptunian belt and if that event were produced under current
conditions, the number of required impacts per year that would be needed to produce
rings as observed, $n_{CHTN}$, may be estimated as

\begin{equation}
n_{CHTN} \approx \pi A^2 P_I C (R_1^{-q} - R_2^{-q})  = 1.3
\times 10^{-12}\ yr^{-1},
\end{equation}

where $P_I= 1.29 10^{-22}\ km^{-2}\ yr^{-1}$ is the intrinsic
probability of collision of trans-Neptunian objects in present-day
conditions \citep{Delloro2013},  
$R_1=0.2\ km$ and $R_2=1.6\ km$ are the minimum and maximum radii of
impactors that can transfer just enough mass to the observed location,
as previously determined. The size distribution of TNOs at small sizes
is fairly unknown, therefore we choose a typical steady-state size
distribution corresponding to $q=3,5$
\citep{Dohnanyi1969,OBrien2005,CampoBagatinBenavidez2012},
%
which should
be a good approximation to the real situation unless any wavy
behavior in the distribution at small sizes is present
\citep{CampoBagatinetal1994}, %
and $C=4.7 10^{4}$ is a
normalizing  constant corresponding to the number of objects
larger than $1\ km$ \citep{Petit2008}.

We also note that the mean
relative velocity in the trans-Neptunian region $v_r(TN) = 1.65\ 
km\cdot s^{-1}$ \citep{Delloro2013} is similar to the required
velocities computed above.

Therefore, we conclude  that it is highly unlikely that this type of
event  
occurred in present-day conditions in the trans-Neptunian belt, during
the age of the solar system.

\section{Catastrophic impact on a previously existent satellite of (10199) Chariklo}

The total mass of the ring has been estimated to be similar to
that of a kilometer-sized object of density $1\ g\cdot cm^{-3}$
\citep{Braga2014}. In this section we therefore calculate the likelihood that the rings of
(10199) Chariklo originated in a catastrophic impact between a field
object and a satellite of the centaur of radius $a_s = 1\ km$.  

According to \citet{Benz1999} the mass of the largest remnant $M_{LR}$
after a catastrophic collision with (10199) Chariklo is related to 
the the incident kinetic energy $K_i$ per unit mass of the target as

\begin{equation}
\frac {M_{LR}}{ M_{Sat}} = -s \left( \frac{K_i}{Q^*} - 1  \right) +
\frac{1}{2}, 
\label{eq:mlr}
\end{equation}

where $M_{Sat}$ is the mass of the centaur's satellite, $s=0.6$ for
ice for $3\ km/s$ impacts. We consider an analytical dependence of
$Q^*$ with the target radius, $R_{Sat}$, as given by \citet{Benz1999}%
, which is valid in both the gravity and the
strength regimes, 

\begin{equation}
Q^*(R_{Sat}) = Q_0\ \left( \frac{R_{Sat}}{1cm} \right)^a + B\ \rho_s\
\left (\frac{R_{Sat}}{1cm}  \right)^b.  
\label{eq:qest}
\end{equation}

Constants for icy targets at $3\ km/s$ are --$Q_0=1.0\ 10^7\ erg/g$, $B=1.2 erg\cdot cm^3 /g$, $a=-0.39$ and $b=1.26$
and we consider only the case of a satellite density $\rho_s = 1 g\cdot
cm^{-3}$. In the range of sizes involved, this expression coincides with that obtained by numerical
simulations of asteroid collisions, which give a value of $b=1.3$
\citep{LeinhardtStewart2009}, and yield similar results. 

The mass of the impactor is taken as $$m_i = \frac{4}{3} \pi \rho_i A_i^3,$$ 
where $A_i$ is its mean radius and $\rho_i$ its
bulk density. The threshold catastrophic collision is defined
when $\frac {M_{LR}}{M_{Sat}} = \frac{1}{2}$. From this, we can
estimate that 
the radius $A_i$ of a projectile capable of disrupting the satellite
of mass $M_{Sat}$ --in an impact at a relative velocity 
$v_i= 3\ km/s$ -- is:

\begin{equation}
A_i =  \left( \frac {3}{ 2 \pi \rho_i \gamma_i} \frac{Q^*(R_{Sat})
M_{Sat}}{v_i^2} \right)^{1/3}.
\label{eq:ai}
\end{equation}

In Figure \ref{fig:Rimp} %
, we plot the radius of the impactor
that produces a catastrophic collision as a function of the radius of
the target, Chariklo's satellite in our case. 
The value of the impact velocity is approximated to be the sum
of the typical
orbital relative velocity and the escape velocity from the
centaur, $v_i= v_{esc} + 3\ km/s$.

We estimate that the radius of an impactor that will catastrophically
disrupt a $3\ km$ satellite is  $A_i \sim 200 m$, if the
bulk density of the target and the impactor are $1\ 
g\cdot cm^{-3}$.

Since the collisional cross section of the satellite is
approximately $10^{-4}$ times that of (10199) Chariklo, the estimated number of
collisions with a satellite would be 
$$n_{Sat} \approx 10^{-12}\ yr^{-1},$$ 
making this event extremely unlikely in the region where
the centaur's orbit is at present. On the other hand, if we estimate the
number of such events, $n_{Sat}$, in the present-day trans-Neptunian
belt, we obtain, 
$$n_{Sat} \approx \pi A_{Sat}^2 P_I C A_i^{-q} = 3.2\times 10^{-14}\ yr^{-1},$$ 
where $A_{Sat} = 3\ km$ is the radius of the
satellite and $A_i = 200 m$, which implies that this type of
collision is extremely rare in the current trans-Neptunian
belt.

\section{Discussion}

In the context of the remarkable discovery of a ring system about the centaur 
(10199) Chariklo, the first question that arises is whether the rings lie inside
or outside the corresponding classical Roche limit. We find that locating
the Roche limit at about 3.3 centaur radii, where the
rings are observed, implies rather extreme physical properties for the
disrupted satellite: 
its density woul have to be $1/3$  that of the primary.

We considered various realistic models for the disruption distance of
an object about the centaur that lead us to conclude that an icy body
with a radius of about $6.5\ km$ or larger \citep{HolsapleMichel2008,
Dobrovolskis1990}, would disaggregate at the ring location. The
minimum amount of mass available after the disaggregation of a
satellite enough to explain: a) the rings, b) the shepherd satellites
capable of confining the system and c) an object which would clear the
observed gap, corresponds to an object of approximately $4\ km$
radius. Therefore a very large amount of mass would have been lost
from the system. If the loose material originated after the disruption
did not fell between the shepherds, it would be swept away as a result
of Pointing-Robertson or viscous drag in a timescale of $10^3 yr$ to
$10^4 yr$ \citep{Braga2014}.  


%


Even if both collision scenarios studied in Sec. 5 and 6 
are physically plausible,
 the estimated timescales are longer than the dynamical lifetime of
centaurs, assuming typical impact probabilities taken form the
literature (see for example \citep[see for example][]{Levison2000}.

If the presently observed rotation period of Chariklo $\sim 7\ hr$ is
primordial, the tidal interaction would have produced the expansion of the
orbit.  The corotation period at the current ring location is $\sim
20\ hr$ and --even if the centaur spin were slower than that, tidal
evolution would be too slow to produce the approach of a small
satellite to its breakup distance. 
By means of a very simple model, we showed that a three-body encounter can be an
effective mechanism to bring a field object of radius $\sim 300 m$ to
its splitting distance, where the rings are observed today. It must be
noted that using Equation \ref{eq:mlr}, the 
collision assumed in our simple model is perfectly ineastic and
results in the merger of the two bodies.

%
%

%
%
%

%


\section{Conclusions}


We estimate that the occurrence of a cratering collision on (10199)
Chariklo or a catastrophic collision on a small satellite are very
rare in the lifetime of a centaur. A remarkable change on our
understanding of the trans-Neptunian population is needed such that
these events become plausible. For instance, the flux of incoming
objects from the trans-Neptunian region should be an order of
magnitude larger than what currently estimated [e.g.
\cite{Hueso2013}]. A ring-forming collision in the primordial more
massive trans-Neptunian belt might be a possible alternative.

If the ring system originated in a tidal disruption, we find that the
parent body should have been approximately $6.5\ km$. The disgregation
of such an
object produces a considerable excess with respect to the currently
estimated mass. We also note that if the
disrupted body were rotating faster than about $5\ hr$ or its shape were
considerably oblated, then the size of the satellite could have been
much smaller \citep{Davidsson2001}, which would  avoid the problem of the excess of mass.

The dynamical mechanism that transported the disrupted object to
that distance was most probably not tidal and it might have beeen associated
with  a multiple encounter, although more detailed modeling is necessary
on this specific issue.

Regarding the composition of the rings, if the material originated
from a collision on the body of (10199) Chariklo, then they would have
came from its outer crust and would most probably be composed of water ice
\citep{Hyodo2016}. as observed \citep{Duffard2014}. 
If the material of the rings comes form a disrupted satellite -either
tidally or collisionally, then we must conclude that this body was
composed mainly of water ice. On the other hand, the ring particles are in constant
physical interaction with each other, therefore it is expected that
their surface are constantly rejuvenated and so, the effects of cosmic
irradiation are rapidly masked.

\begin{acknowledgements}

RD acknowledge the support of MINECO for his Ram\'on y Cajal Contract.
The research leading to these results has received funding from the European
Union’s Horizon 2020 Research and Innovation Programme, under Grant
Agreement No. 687378. Funding from
Spanish funding form grant
AYA-2014-56637-C2-1-P is acknowledged, as is the Proyecto de
Excelencia de la Junta de Andalucía, J. A. 2012-FQM1776. 
FEDER funds are also acknowledged. MDM acknowledge the support of
MinCyT ANPCyT PICT 1144/2013 and the support of the travel grant from
CONICET and CSIC.

\end{acknowledgements}

\bibliographystyle{aa} 
\bibliography{RingsChariklo}

\clearpage

\newpage

\clearpage

\begin{figure}[p!]
\centering
\caption{Splitting distance from (10199) Chariklo as a function of the
satellite radius for two different strength regimes under the
assumptions of \citet{Dobrovolskis1990}. }
\label{fig:3}
\end{figure}

\clearpage

\begin{figure}[p!]
\centering
\includegraphics[angle=-90,width=\textwidth]{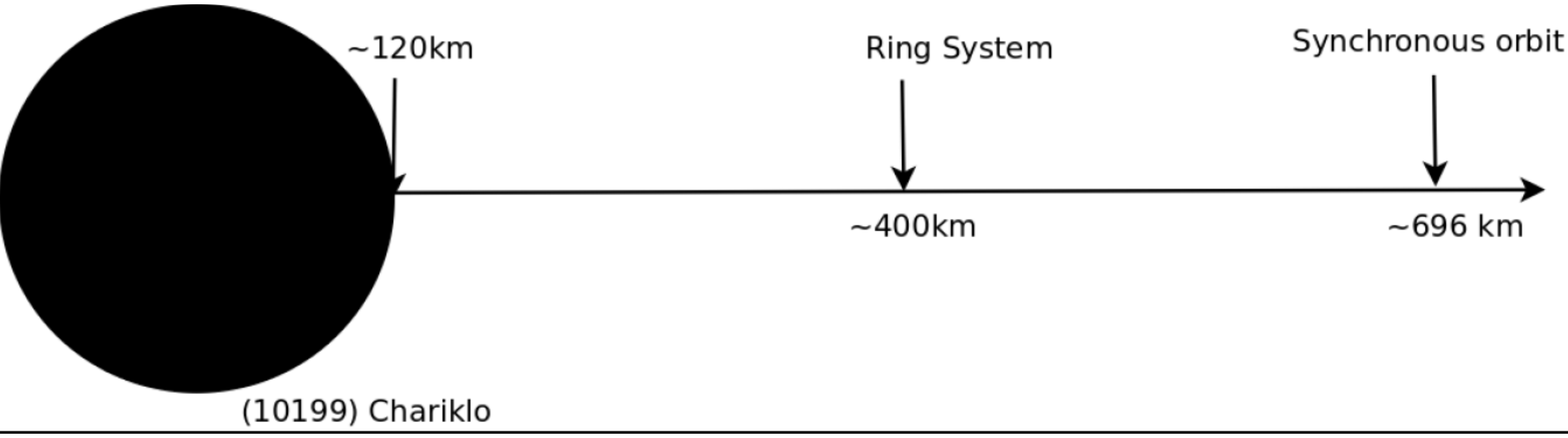}
\caption{Relative location of the synchronous orbit and the location of
the rings. 
}
\label{fig:sk0}
\end{figure}

\clearpage

\begin{figure}[p!]
\centering
\includegraphics[scale=0.6,angle=270]{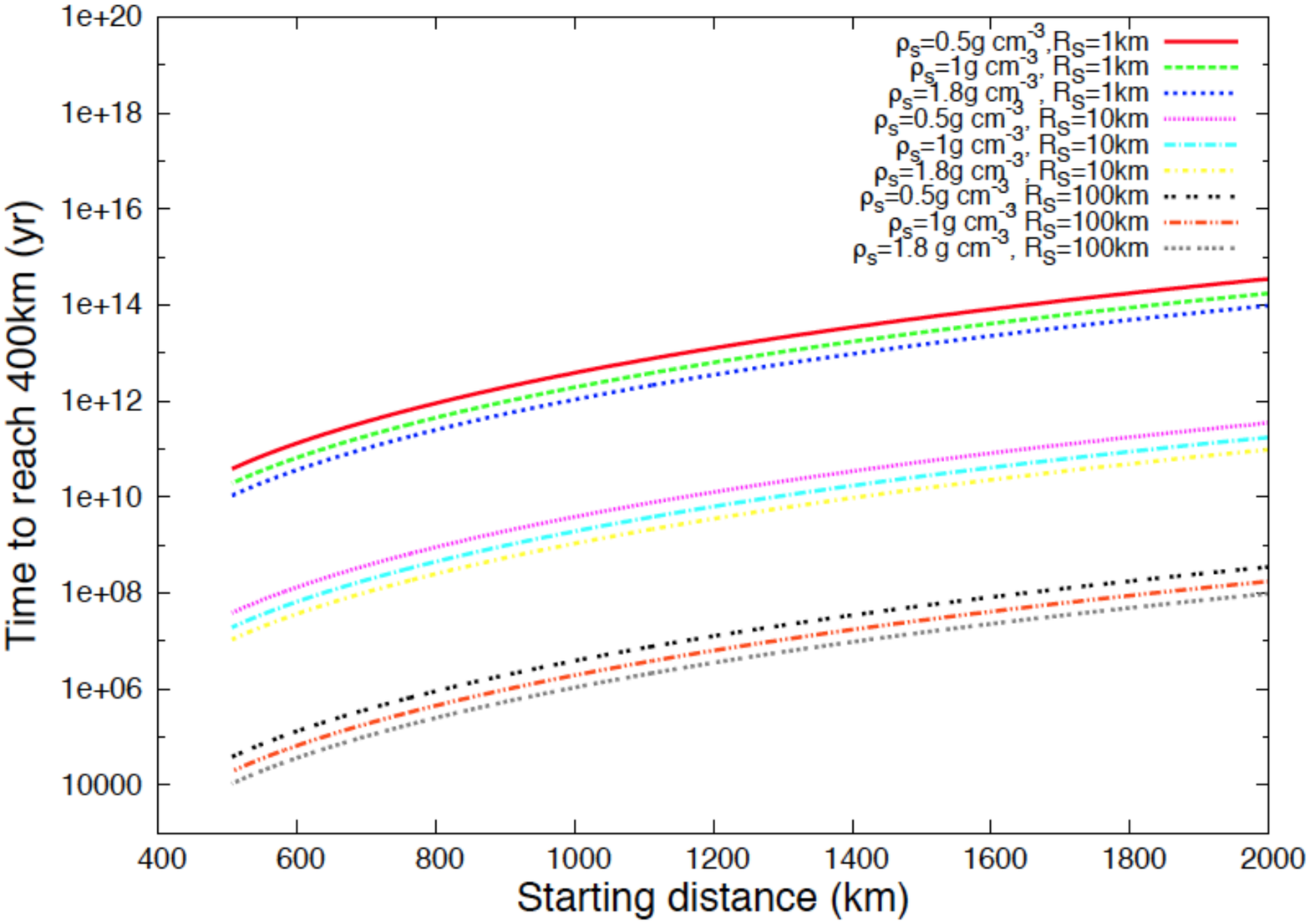}
\caption{
Tidal-evolution travel timescale, $\tau_{1 2}$, due to the
tides raised on the centaur, as a function of the starting
distance assumed to be larger than $500\ km$, for different values
of the radius of the satellite and its density. }
\label{fig:5}
\end{figure}

\clearpage

\begin{figure}[p!]
\centering
\includegraphics[scale=0.6,angle=270]{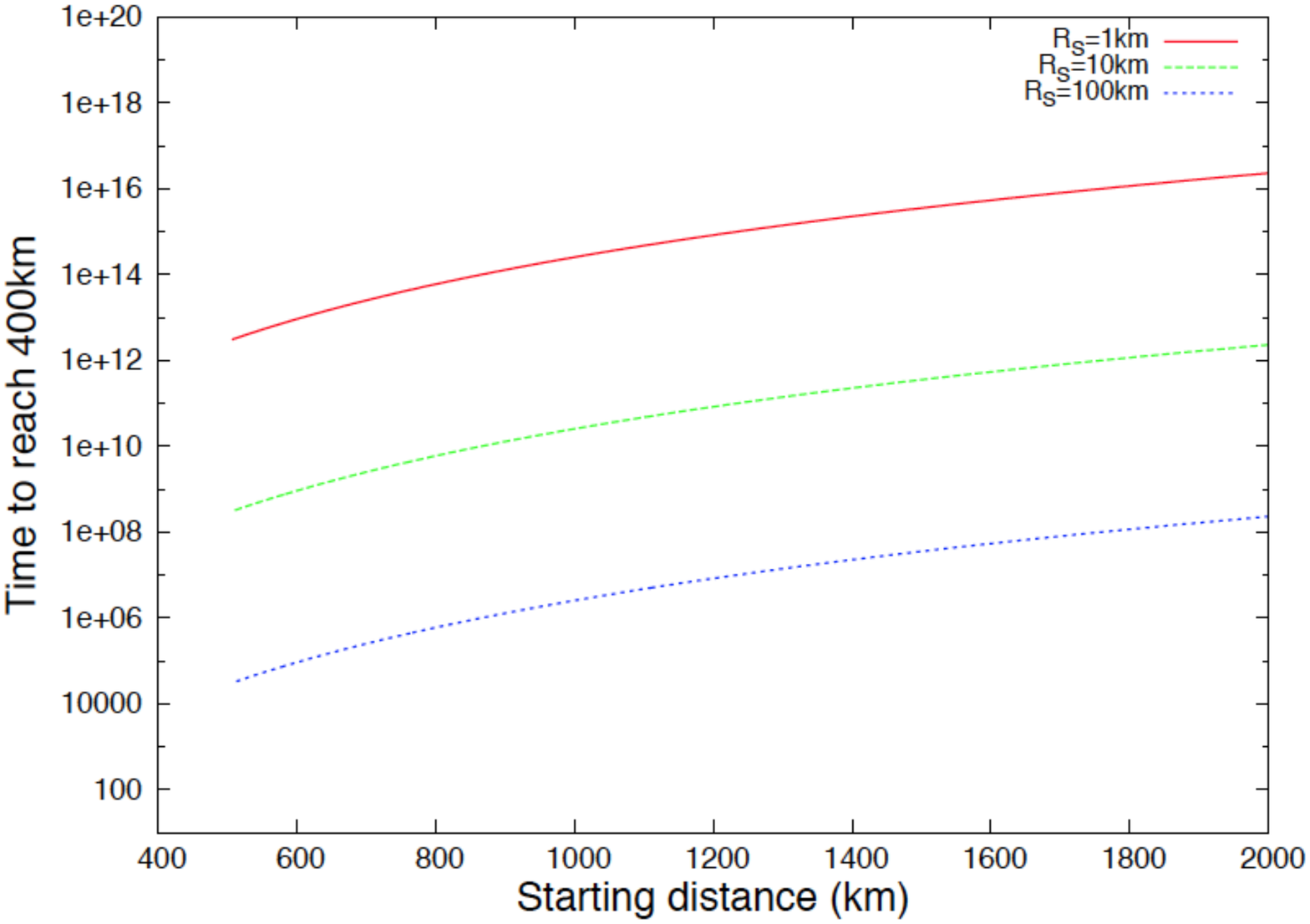}
\caption{
Tidal-evolution travel timescale, $\tau'_{1 2}$, due to the
tides raised on the satellite, as a function of the starting distance
assumed to be larger than $500\ km$, for different values
of the radius of the satellite. We also assume that the density of the
satellite is $1 g\cdot cm^{-3}$.}
\label{fig:5b}
\end{figure}

\clearpage

\begin{figure}[p!]
\centering \includegraphics[scale=0.6,angle=270]{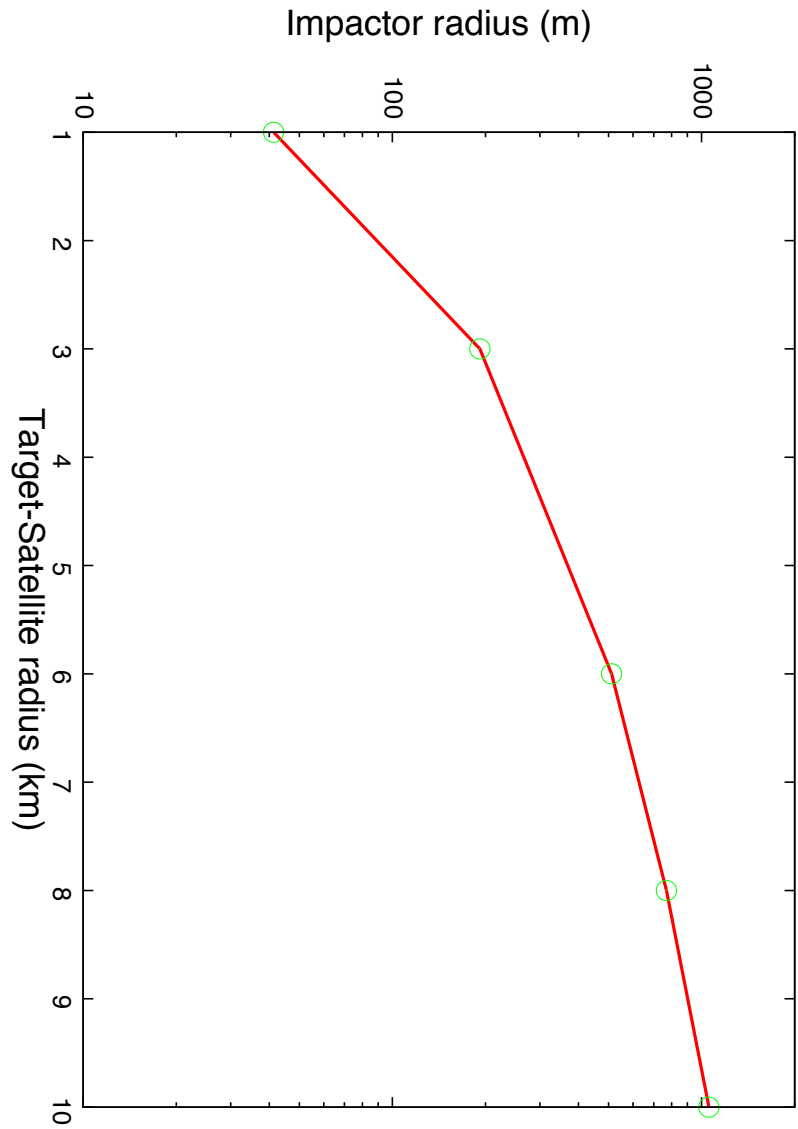} \caption{
Radius of the impactor that produces a shattering collision at a
relative velocity of $3\ km\cdot s^{-1}$, as a function of target
radius. We assume that the density of both the impactor and the
target satellite is $1 g\cdot cm^{-3}$.} \label{fig:Rimp}
\end{figure}

\clearpage

\begin{figure}[p!]
\centering
\includegraphics[width=\textwidth]{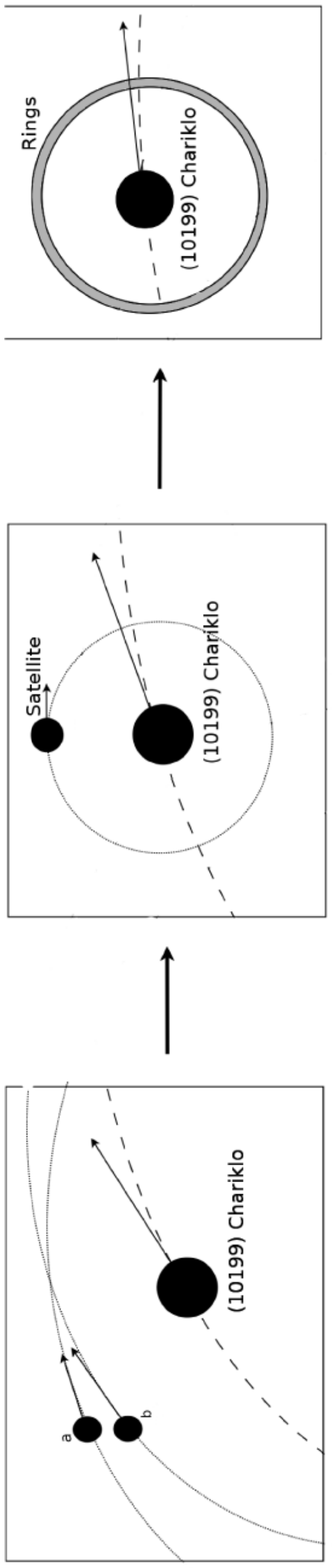}
\caption{Three body encounter sketch. A perfectly inelastic collision
between field objects $a$ and $b$ dissipate enough energy such that the
combined object end up in orbit around the centaur. }
\label{fig:sk1}
\end{figure}

\clearpage

\begin{figure}[p!]
\centering
\includegraphics[scale=0.9]{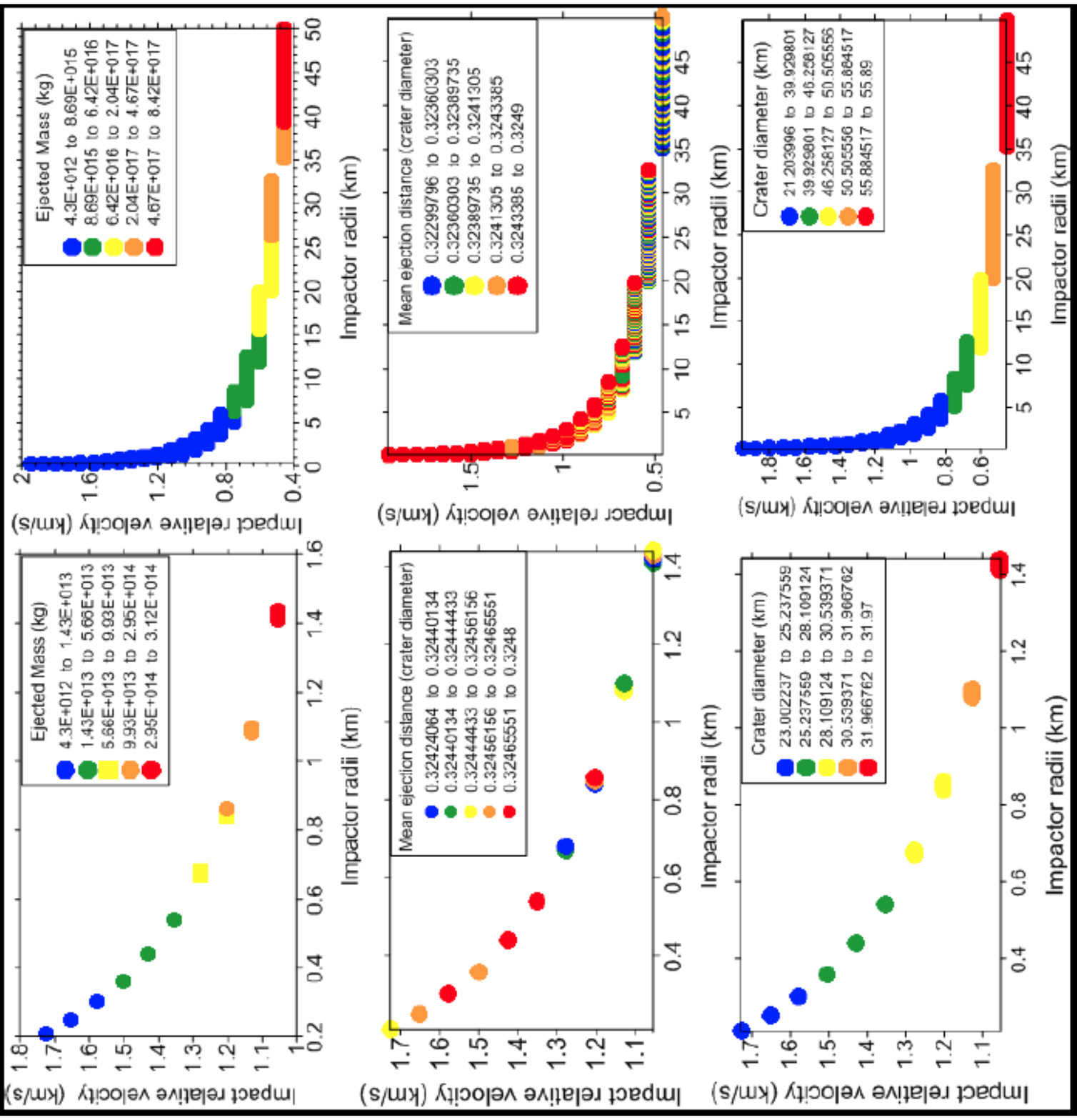}
\caption{Classed post plots of ejected mass, against impact relative
velocity and size of the impactor. The constraint exists is that 
that the ejection velocities are in the range given by condition \ref{c1}. }
\label{fig:vt}
\end{figure}

\end{document}